\documentclass[final]{jfm}

\usepackage{commath}

\usepackage[hidelinks]{hyperref}
\usepackage[nameinlink,noabbrev]{cleveref}%
\crefname{figure}{Figure}{Figures}
\crefname{equation}{Equation}{Equations}

\usepackage{graphicx}
\usepackage[numbers]{natbib}

\usepackage[locale=UK,
number-mode=math,
unit-mode=text,
per-mode=symbol,
number-unit-product=\ ,
retain-zero-exponent=true]{siunitx}
\DeclareSIUnit{\pixel}{px}
\DeclareSIUnit{\fps}{fps}

\newcommand{\kindex}[2]{\ensuremath{{#1}_{\scalebox{0.5}{#2}}}}

\newcommand\dd{\textrm{d}}

\graphicspath{{../../figurematter/latex/figs/}}

\begin{document}
\title{Discrete shedding of secondary vortices along a modified Kaden spiral}

\author{Diego Francescangeli, Karen Mulleners$^{*}$}
\affiliation{Institute of Mechanical Engineering, \'Ecole polytechnique f\'ed\'erale de Lausanne, Switzerland\\
$^*$ Corresponding author: karen.mulleners@epfl.ch
}

\maketitle
\begin{abstract}
When an object is accelerated in a fluid, a primary vortex is formed through the roll-up of a shear layer.
This primary vortex does not grow indefinitely and will reach a limiting size and strength.
Additional vorticity beyond the critical limit will end up in a trailing shear layer and accumulate into secondary vortices.
The secondary vortices are typically considerably smaller than the primary vortex.
In this paper, we focus on the formation, shedding, and trajectory of secondary vortices generated by a rotating rectangular plate in a quiescent fluid using time-resolved particle image velocimetry.
The Reynolds number \Rey based on the maximum rotational velocity of the plate and the distance between the centre of rotation and the tip of the plate is varied from \numrange{840}{11150}.
At low \Rey, the shear layer is a continuous uninterrupted layer of vorticity that rolls up into a single coherent primary vortex.
At $\Rey = \num{1955}$, the shear layer becomes unstable and secondary vortices emerge and subsequently move away from the tip of the plate.
For $\Rey > 4000$, secondary vortices are discretely released from the plate tip and are not generated from the stretching of an unstable shear layer.
First, we demonstrate that the roll-up of the shear layer, the trajectory of the primary vortex, and the path of secondary vortices can be predicted by a modified Kaden spiral for the entire \Rey range considered.
Second, the timing of the secondary vortex shedding is analysed using the swirling strength criterion.
The separation time of each secondary vortex is identified as a local maximum in the temporal evolution of the average swirling strength close to the plate tip.
The time interval between the release of successive secondary vortices is not constant during the rotation but increases the more vortices have been shed.
The shedding time interval also increases with decreasing Reynolds number.
The increased time interval under both conditions is due to a reduced circulation feeding rate.
\end{abstract}

\section{Introduction}
The life of vortices around bluff bodies often begins with a shear layer \citep{fernandoSeparationMechanicsAccelerating2017, jeonRelationshipVortexFormation2004a, rosiEntrainmentTopologyAccelerating2017, fernandoVortexEvolutionWake2016, corkeryQuantificationAddedmassEffects2019}.
When a bluff body moves relative to a fluid flow, a thin layer of fluid emerges at the edge of the body where non-zero shear flow gradients are present.
This shear layer is characterised by increased values of the flow vorticity.
In the wake of the body, the shear layer rolls-up and the shear layer vorticity accumulates into a coherent vortex.
The interplay between the free stream or body's velocity and the induced velocity of the growing coherent vortex cause the shear layer to become curved.
This curvature changes continuously in time.
The roll-up of a semi-infinite shear layer or vortex sheet was first described by \citet{kadenAufwicklungUnstabilenUnstetigkeitsflaeche1931}, who derived the following self-similar equation to describe the shear layer shape at any point in time $t$:
\begin{equation}
r = K(t/ \theta )^{2/3}
\label{Kaden}
\end{equation}
where $K$ is a dimensional constant, and $r$ and $\theta$ are the radial and angular coordinates along the spiral with $r = 0, \theta\rightarrow\infty$ at the spiral centre, $r\rightarrow\infty, \theta\rightarrow 0 $ at the opposite end of the semi-sheet at infinity.
The exponent $2/3$ is retrieved from dimensional analysis and the obtained curve is a spiral with tight inner turns.
The initial strength of the flat sheet increases monotonically with increasing distance away from the tip of the body.
For $t>0$, the spiral has an infinite number of turns leading to a singularity of the velocity and the sheet strength decreases to zero for $\theta \to \infty$ in the spiral centre.
The maximum value of the sheet strength is now located somewhere along the sheet \citep{saffmanVortexDynamics1995}.
In reality, viscosity will remove any singularity at the spiral centre and yield the development of a viscous core~\citep{mooreAxialFlowLaminar1973}.

At the early stages of the roll-up, Kaden's spiral is tight with a low local radius of curvature.
It accurately represents the initial evolution of the shear layer.
At later stages, the radius of curvature increases due to the viscous interactions within the shear layer and between the shear layer and the coherent primary vortex that grows due to the continuous accumulation of vorticity at the centre of the spiral.
The distortions can be investigated by modelling the inner portion of the spiral as a single point vortex located at the centre~\citep{mooreNumericalStudyRollup1974}.
The entire shear layer roll-up can also be predicted by a point-vortex representation of an initially straight vortex sheet \citep{krasnyComputationVortexSheet1987, devoriaVortexSheetRollup2018}.
The degree of the elliptical distortions depends on the shape of the object.
They are almost negligible for flat plates and become more pronounced when the edge has a non-zero wedge angle~\citep{pullinLargescaleStructureUnsteady1978}.

The accumulation of the vorticity in the coherent vortex in the spiral centre does not continue indefinitely.
There is a physical limit to the size and the amount of circulation the primary vortex can collect \citep{gharibUniversalTimeScale1998, Mohseni1998, Gao:2010ia, deguyon2020universal}.
When the primary vortex is about to pinch-off, a trailing pressure maximum is observed along the shear layer \citep{lawsonFormationTurbulentVortex2013}.
The shear layer region between the tip and the trailing pressure maximum has an adverse pressure gradient.
The remaining portion of the shear layer is characterised by a positive pressure gradient.
The two regions of the shear layer are now separated and the vorticity associated with the adverse pressure gradient can not be entrained into the vortex core.
The trailing pressure maximum travels downstream together with the primary vortex, causing the subsequent pinch-off of the primary vortex \citep{SchlueterKuck:2016kw}.
Additional vorticity will not be entrained by the primary vortex after pinch-off and instead can accumulate into smaller secondary vortices within the trailing shear layer similar to a Kelvin-Helmholtz instability \citep{dabiriOptimalVortexFormation2009}.
The increases in shear layer curvature during the initial stages of the vortex formation momentarily stops when the end of the primary vortex growth is reached \citep{sattariGrowthSeparationStartup2012}.
Secondary vortices occur first between the primary vortex and the tip at locations where the sheet strength according to \citeauthor{kadenAufwicklungUnstabilenUnstetigkeitsflaeche1931} is maximal \citep{mooreNumericalStudyRollup1974, koumoutsakosSimulationsViscousFlow1996}.

The emergence of secondary vortices seem to occur only if the Reynolds number is above a critical threshold.
The value of this critical Reynolds number varies for different object geometries and boundary conditions.
Critical values in a range from $\Rey = $\numrange{1000}{3000} were observed in a cylinder wake by \cite{wuShearLayerVortices1996a}.
The lower limit was slightly higher for \cite{bloorTransitionTurbulenceWake1964}, who did not detect any instabilities for $\Rey < 1300$.
The span-wise and end configurations strongly affect the shear layer breaking behind a cylinder.
Parallel and oblique vortex shedding are obtained by changing the inclination of end plates \citep{prasadInstabilityShearLayer1997}.
The shear layer manifests instabilities at $\Rey = 1200$ for parallel shedding and  at $\Rey = 2600$ for oblique shedding.
The critical Reynolds number for an accelerated sharp edged plate lies in a higher range.
\citet{pullinFlowVisualizationExperiments1980,williamsonVortexDynamicsCylinder1996} started to visually observed secondary vortices along the shear layer for $\Rey = 4268$.
This value was later confirmed by \citet{luchiniStartupVortexIssuing2002}, who numerically observed the occurrence of secondary vortices in a range from $\Rey =$ \numrange{4500}{5000}.

For Reynolds numbers above the critical value, series of secondary vortices appear in the trailing shear layer with a seemingly constant distance between them.
The typical shear layer frequency in the wakes of cylinders is much higher than the frequency of the von Karman vortex street.
A consensus about the exact relationship between the frequency of the primary vortex shedding $\kindex{f}{K}$ and the secondary of shear layer vortices  $\kindex{f}{SL}$ has not yet been found.
\citet{bloorTransitionTurbulenceWake1964} observed that the ratio between the characteristic frequencies varies with Reynolds number according to $\kindex{f}{K}/\kindex{f}{SL} = \Rey^{1/2}$.
However, there is no consensus about the exponent value of the proposed relationship.
\citet{prasadInstabilityShearLayer1997} indicated that an exponent value of $0.67$ works for $\Rey$ up to $10^5$ and \citet{weiSecondaryVorticesWake1986} found $0.87$ in the range from $\Rey =$ \numrange{1200}{11000}.
No clear relationships are established in the situation of an isolated primary vortex.
Based on the flow visualisation around a submerged flat plate, \citet{griftDragForceAccelerating2019} determined the shedding frequency of secondary vortices to lie in the range from \SIrange{13}{20}{\hertz}, for different values of acceleration, velocity, and immersion depth.
This range corresponds to a Strouhal number around $0.2$, according to the plate geometry and kinematics used by the authors.
The secondary vortex shedding frequency behind a vertical flat plate increases with increasing acceleration of the flat plate according to \citet{rosiEntrainmentTopologyAccelerating2017}.
It is crucial to define a scaling parameter, such as the Strouhal frequency for the cylinder case, that allows for a more universal relationship between the shedding frequency or formation time of primary and secondary vortices as a function of the Reynolds number.

Secondary vortices also have a practical relevance for a broad range of applications.
They can create additional lift on delta wings at high angles of attack \citep{gad-el-hakDiscreteVorticesDelta1985}, cause vortex induced oscillations of solid structure that lead to fatigue damage \citep{shiraishiClassificationVortexinducedOscillation1983}, and lead to increased drag and noise for wing tip vortices \citep{Birch2003}.
A precise prediction of secondary vortices can improve aerodynamic performance and reduce vortex induced vibrations and noise.

Here, we present an experimental study of secondary vortices generated by a rotating flat plate in a quiescent fluid.
The experimental setup is discussed in details in the following section and is similar to the configurations used by \citet{davidKinematicGenesisVortex2018, corkeryQuantificationAddedmassEffects2019, carrAspectratioEffectsRotating2015a}.
The plate is rotated with a constant rotational velocity which is varied across different experiments.
The rotation of the plate generates a start up or a primary vortex.
As the plate keeps rotating, the primary vortex separates and smaller secondary vortices are observed.
First, we determine the critical Reynolds number above which secondary vortices are observed in the shear layer behind the tip of the rotating plate.
Second, we describe the path of secondary vortices and model their path using a modified Kaden spiral.
Finally, we estimate the timing of the secondary vortex shedding process and analyse the effect of the Reynolds number on the timing.

\section{Experimental methods}
\begin{figure}
\centerline{\includegraphics{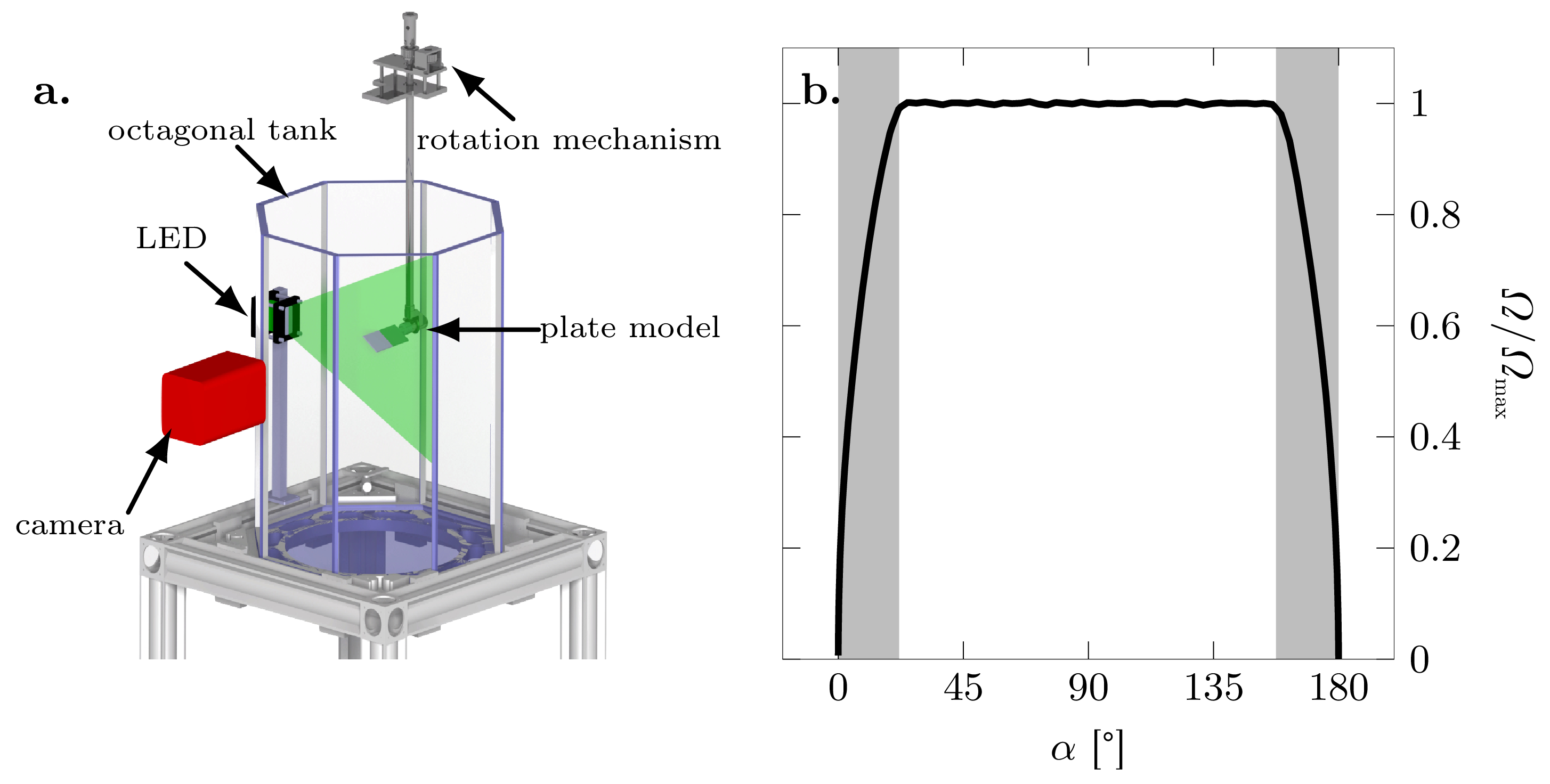}}
\caption{
(a) Schematic of the experimental set-up and the rotation mechanism.
(b) Trapezoidal velocity profile as a function of the angular position.
The grey shaded regions indicate the portion of the motion during which the plate is accelerated.
}
\label{fig:setup}
\end{figure}

The first series of measurements is conducted with a rectangular flat glass plate, with length $l=\SI{8}{\centi\meter}$, width or span $s=\SI{16}{\centi\meter}$ and thickness $t=\SI{2}{\milli\meter}$ that is rotated about \ang{180} in a water tank around its centre span-wise axis.
The distance between the centre of rotation and the tip of the plate is referred to as the chord length $c$ here.
The length of the plate is reduced to $l=\SI{4}{\centi\meter}$ and the rotation point is shifted to the edge of the plate for the second set of measurements.
The chord length or distance between the rotational point and the tip of the plate is preserved for both sets of experiments.
For the first set of experiments, vortices are formed symmetrically behind both ends of the plate.
For the second set of experiments, vortices are formed only on one end of the plate.
This allows us to study the influence of the rotation point and detect potential interferences caused by symmetric vortex release on both tips when the rotation point is at mid-length.
A third set of measurements with a longer plate with length $l=\SI{12}{\centi\meter}$ and the rotation point at mid-length, yielding a chord length of $c=\SI{6}{cm}$, was conducted to provide insight into the influence of the chord length on the vortex formation.
The glass plate is stiff enough to not bend due to the interaction with water and its transparency prevents shadow regions when performing particle image velocimetry (PIV).
The experiments are conducted in an octagonal tank with an outer diameter of $\SI{0.75}{\meter}$ filled with water (\cref{fig:setup}a).

The rotation mechanism is fastened to an outer aluminium frame such that the mid span of the plate is in the centre of the tank to limit wall interference effects.
The kinematic input is given by a servo motor (Maxon RE $35$) connected to a stainless steel shaft and transferred to the flat plate through a $1:1$ conical coupling.
A $1:19$ gearbox is mounted on the motor to ensure high torque, speed, and acceleration.
The rotational angle, speed, and acceleration are controlled via a Galil DMC-40 motion controller, which allows for accurate control of arbitrary motion profiles.
The rotation programme is a trapezoidal rotational velocity profile with a fixed rotational amplitude of $\ang{180}$ (\cref{fig:setup}b).

To ensure a continuous acceleration profile, the corners of the velocity trapezoid are smoothed.
The maximum rotational speed $\kindex{\Omega}{max}$ is varied from \SIrange{30}{400}{\degree\per\second}.
This leads to a Reynolds number $\Rey = (\kindex{\Omega}{max}c^2)/\nu$ based on the chord, which is defined as the distance between the rotation point and the tip of the plate, ranging from \numrange{840}{11150}.
The rotational acceleration $\dot{\Omega}$ is fixed at $\SI{6000}{\degree\per\second\squared}$.

The PIV images are recorded in the cross-sectional plane at the model mid span.
A high-power pulsed light-emitting diode (LED Pulsed System, ILA $5150$ GmbH) is used to create a light sheet in the measurement plane.
The applicability of high-power LED for PIV has been demonstrated previously by \citet{willertPulsedOperationHighpower2010, krishnaFlowfieldForceEvolution2018}.
Time-resolved PIV images are recorded with a Photron FASTCAM SA-X$2$ high speed camera.
The camera is equipped with a $\SI{35}{\milli\meter}$ Canon lens and the camera is aligned carefully such that the optical axis of the lens is aligned with the rotational axis of the plate and is perpendicular to the light sheet (\cref{fig:setup}a).
The frame rate and the exposure time are varied, depending on the dynamics of the motion.
A frame rate and exposure time of $\SI{250}{\hertz}$ and $\SI{1}{\milli\second}$ are selected for a rotational speed of $\SI{30}{\degree\per\second}$.
These values are $\SI{2000}{\hertz}$ and $\SI{0.5}{\milli\second}$ for the highest tested speeds.
The frame rate is high enough to capture the dynamics of the motion and the LED is set to continuous mode.
The camera resolution is $\SI{1024 x 1024}{px}$, which corresponds to a field of view of $\SI{20 x 20}{\centi\meter}$.
The raw data are processed by the commercial software PIVview (PIVTEC GmbH, ILA 5150 GmbH) using a correlation model based on minimum squared differences and a multi-pass interrogation algorithm with three iterations.
The final interrogation window size is $\SI{32 x 32}{px}$ with an overlap of $\SI{68}{\percent}$.
A third order B-spline interpolation method for sub-pixel image shifting is performed on all passes.
The resulting physical resolution is $\SI{1}{\milli\meter}$ or $0.025$c.

\section{Results}
\subsection{Modelling the shear layer roll-up}
\begin{figure}
\begin{center}\includegraphics{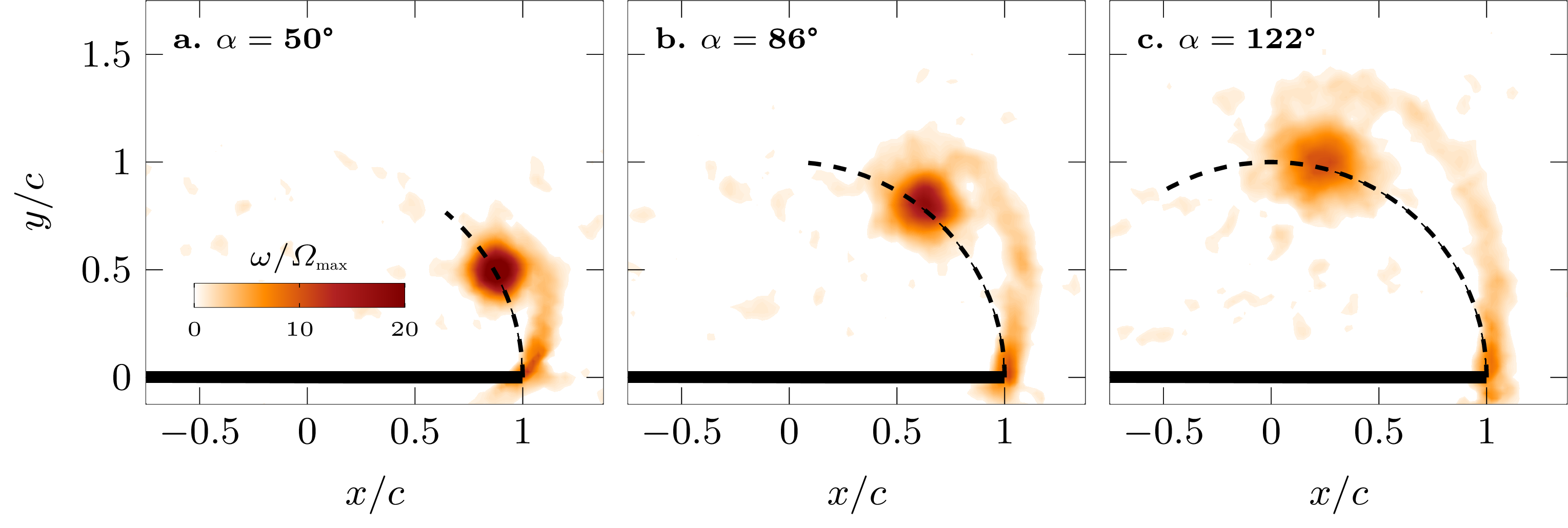}\end{center}
\caption{Vorticity fields for different angular positions (a) $\alpha=\ang{50}$, (b) $\alpha=\ang{86}$, and (c)~$\alpha=\ang{122}$ for $\Rey = \num{840}$.
The dashed line represents the plate tip trajectory.}
\label{fig_continuous_SL}
\end{figure}

At $\Rey= \num{840}$, the plate rotation gives rise to the formation of a primary vortex (\cref{fig_continuous_SL}).
The vorticity fields at different angular positions are shown in the plate's frame of reference.
The primary vortex is the only coherent structure that can be observed and it is connected to the plate tip through a continuous shear layer.
No sign of instabilities are observed in the shear layer as the plate continues the rotation.
The shear layer remains connected to the primary vortex and rolls-up around its core.
As a consequence, the shear layer roll-ups into a spiral that continuously grows in time.

\begin{figure}
\begin{center}\includegraphics{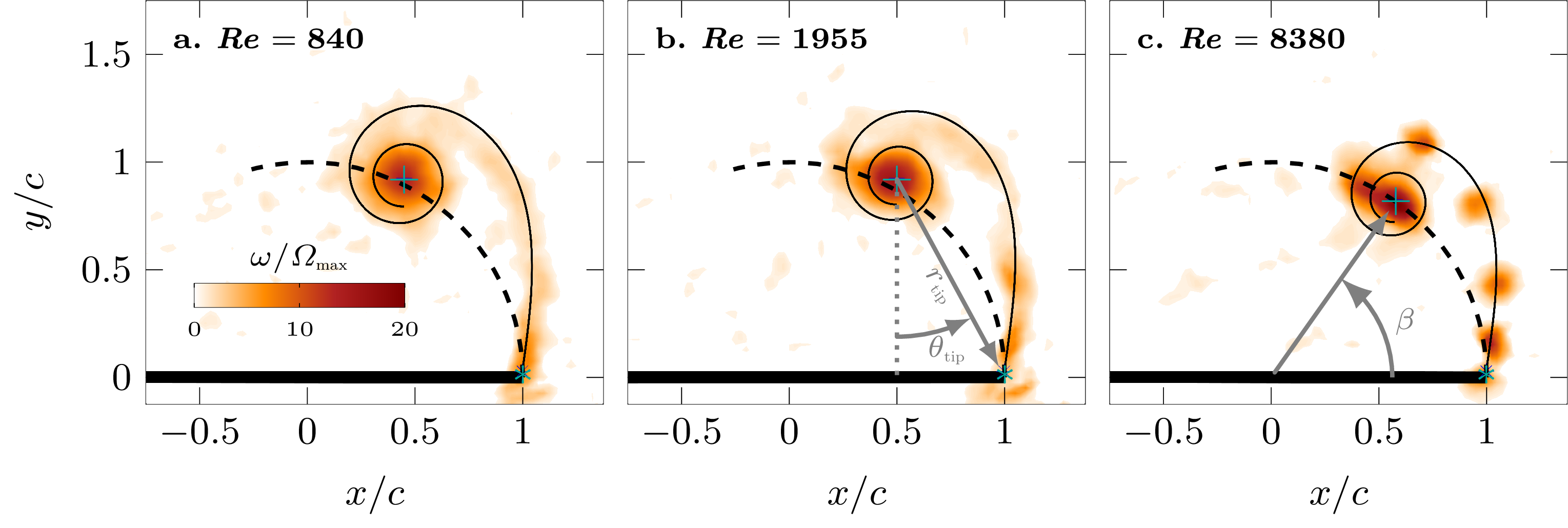}\end{center}
\caption{
Fit of the Kaden's spiral (black solid curve) atop of instantaneous vorticity fields at $\alpha=\ang{105}$ for (a) $\Rey = \num{840}$, (b) $\Rey = \num{1955}$, and (c) $\Rey = \num{8380}$.
The marker $\ast$ indicates the top right edge of the plate and the point where the spiral ends, $+$ indicates the centre of the primary vortex and the point where the spiral begins.
The spiral is only plotted for $\theta$ ranging from \kindex{\theta}{tip} to $4\pi$.
The dashed line represents the plate tip trajectory.
}
\label{fig_Kadencomparison}
\end{figure}

To trace the spiralling topology of the shear layer in the individual snapshots, we start by fitting the Kaden spiral (\cref{Kaden}) to the experimental data.
At every time instant, the Kaden parameter $K$ is determined such that the spiral passes through the plate's edge when the spiral centre is shifted to the instantaneous location of the primary vortex core.
The location of the primary vortex core was retrieved using the $\kindex{\Gamma}{2}$ criterion by \citet{graftieauxCombiningPIVPOD2001}.
The resulting Kaden spirals are presented in \cref{fig_Kadencomparison} atop three instantaneous vorticity snapshots after a rotation of $\alpha=\ang{105}$ for increasing values of the Reynolds number: $\Rey={840,1955,8380}$.
The dashed lines in \cref{fig_Kadencomparison} indicate the plate tip trajectory since the start of the motion, the markers indicate the centre location of the primary vortex, and the solid lines are the fitted Kaden spirals.
For all three Reynolds numbers, the centre of the primary vortex is located on the plate tip trajectory and the fitted spirals match the rolling up shear layer well based on visual inspection.
The vorticity concentration along the shear layer evolves with increasing Reynolds number from a continuous band of vorticity at $\Rey= \num{840}$ (\cref{fig_Kadencomparison}a) to an alignment of vorticity lumped into discrete vortices at $\Rey=\num{8380}$ (\cref{fig_Kadencomparison}c).
At the intermediate Reynolds number $\Rey=\num{1955}$, the shear layer is undulating and some localised concentrations of high vorticity can be identified along it (\cref{fig_Kadencomparison}b).
These are signs of an unstable shear layer.
When we further increase the Reynolds number to \num{8380}, the shear layer instability becomes more prominent.
The primary vortex is no longer connected to the plate tip and the shear layer is broken into a series of distinct individual structures that we refer to as secondary vortices (\cref{fig_Kadencomparison}c).
The fit of Kaden's spiral still describes well the unstable shear layer evolution and goes through the secondary vortices for the entire range of Reynolds numbers considered here.

\begin{figure}
\begin{center}\includegraphics{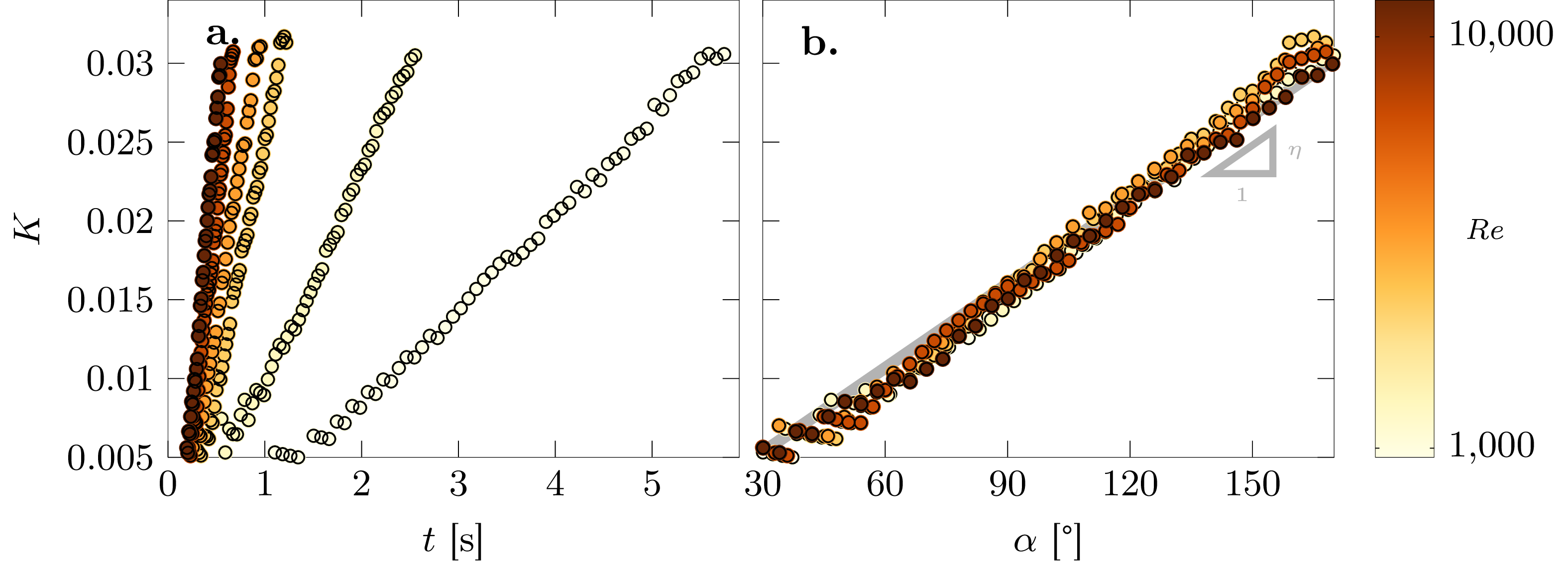}\end{center}
\caption{$K$ parameter of Kaden's equation as a function of (a) time and (b) angular position of the plate for all the tested Reynolds numbers.}
\label{fig:Kconstant}
\end{figure}

So far, we have merely fitted \cref{Kaden} to our experimental data at every time instant, treating Kaden's constant $K$ as a fitting parameter.
We observe that the main topology of the roll up is well captured by the Kaden spiral, but we have not yet gained any insight into the temporal evolution of the roll up or the motion of the primary vortex.
If our shear layer would follow the time evolution predicted by Kaden's spiral, the obtained values for $K$ should be constant for all time instants.
Based on the results presented in \cref{fig:Kconstant}, we conclude that $K$ is not a constant value for our data but increases linearly in time for all Reynolds numbers.
The rate of increase of $K$ with dimensional time decreases with increasing \Rey (\cref{fig:Kconstant}a), but all curves collapse when presented in terms of the angular position of the plate (\cref{fig:Kconstant}b).
The angular position of the plate serves as the dimensionless time variable.
It corresponds to the ratio between the travelled arc length $l = \Omega t c$ and the chord length and represents a convective time scale.
The chord length refers to the length between the centre of rotation and the tip of the plate.

Based on these results, we propose here a modified version of the Kaden spiral to describe and predict the temporal evolution of the shear layer roll up:
\begin{equation}\label{eq:modKaden}
r=\eta \alpha \left(\frac{\alpha}{\theta}\right)^{2/3},
\end{equation}
where $r$ and $\theta$ are again the radial and angular coordinates of the spiral with respect to the spiral centre or primary vortex centre, $\alpha$ is the angular position of the plate and $\eta \alpha$ replaces the dimensional constant $K$ in Kaden's formulation (\cref{Kaden}).
The value of $\eta$ is constant for all \Rey and is empirically determined based on the ensemble of experimental data to $\eta=\num{1.02e-2}$.
The original solution of the Kaden spiral was derived for an unbound semi-infinite vortex sheet that starts out as a straight vortex sheet~\cite{kadenAufwicklungUnstabilenUnstetigkeitsflaeche1931}.
The open end of the sheet rolls up into a vortex with the centre at $(r,\theta)=(0,\infty)$ and the other side of the vortex sheet is at infinity $(r,\theta)=(\infty, 0)$.
For our experimental conditions, the vortex sheet length is finite and its length increases in time.
The open end rolls up into a primary vortex.
The bound end of the vortex sheet is attached to the tip of the rotating plate and only the portion of the modified Kaden spiral for $\theta\in[\kindex{\theta}{tip}, \infty]$ corresponds to our finite shear layer.
Here, $\kindex{\theta}{tip}$ decreases in time and indicates the bound end that is connected to the plate tip.
The value of $\kindex{\theta}{tip}$ is determined at every time step based solely on the observation that the primary vortex moves along a path that matches the plate tip trajectory as indicated in \cref{fig_continuous_SL} by the dashed line.
Based on this purely geometric constraint, we also directly obtain the radial spiral coordinate where the modified Kaden spiral meets the plate tip, indicated by $\kindex{r}{tip}$, and the angular location of the primary vortex with respect to the plate, denoted by $\beta$.
The detailed derivation of $\kindex{\theta}{tip}$, $\kindex{r}{tip}$, and $\beta$ is provided in \cref{app:modKaden}.
With this additional information, we can now write the spatial coordinates of the spiral in the plate's frame of reference as:
\begin{align}\label{eq:modKadenspiral}
\kindex{x}{spiral}&= r\sin{\theta} + c \cos{\beta}\\
\kindex{y}{spiral}&=-r\cos{\theta} + c \sin{\beta},
\end{align}
with $\theta$ and $\beta$ defined as indicated in \cref{fig_Kadencomparison}.
This modified version of the Kaden spiral is now a fully predictive model of the shear layer roll-up and the position of the primary vortex core with a single empirical constant $\eta$.

\subsection{Validation of the model}
The ability of our modified Kaden spiral to describe the roll-up of the shear layer is visually compared to the negative finite time Lyapunov exponent (nFTLE) fields corresponding to the vorticity fields presented in \cref{fig_Kadencomparison}a-c.
The FTLE is a local measure of Lagrangian stretching of evolving fluid particle trajectories \cite{Haller2001, Haller2015}.
The maximising ridges of the negative FTLE field indicate regions along which nearby fluid particles are attracted such as the boundaries of coherent structures.
The FTLE ridges provide insight into the location and growth of vortices and the flow topology \cite{Green2007, Rockwood:2018cl}.

\begin{figure}
\begin{center}\includegraphics{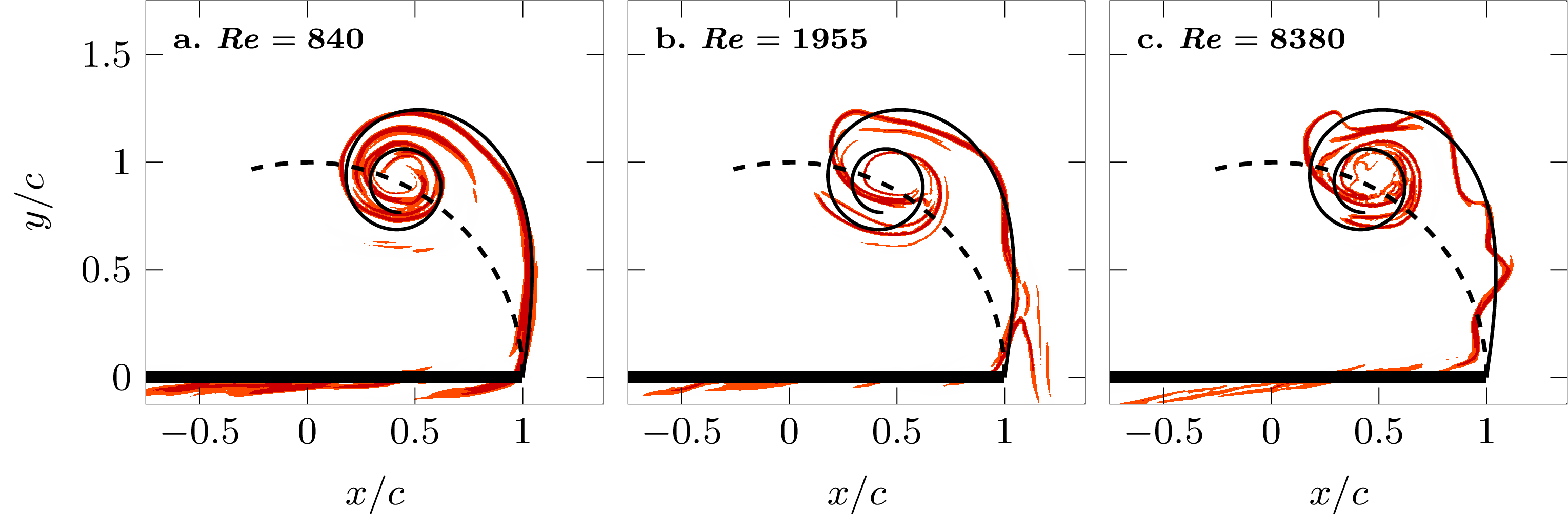}\end{center}
\caption{
Model of the shear layer roll-up (solid curve) atop of nFTLE fields at $\alpha=\ang{105}$ for (a) $\Rey = \num{840}$, (b) $\Rey = \num{1955}$, and (c) $\Rey = \num{8380}$.
The spiral is only plotted for $\theta$ ranging from \kindex{\theta}{tip} to $4\pi$.
}
\label{fig:FTLEregimes}
\end{figure}

At $\Rey = \num{840}$ the shear layer is continuous and the attracting nFTLE ridges appears as a continuous spiral.
The shape and the roll-up of the spiral is well described by our predictive model (\cref{fig:FTLEregimes}a).
At $\Rey = \num{1955}$, we are in a transitional regime where the shear layer is wavy and unstable (\cref{fig_Kadencomparison}b).
This observation is confirmed by the FTLE ridges, where the attracting nFTLE ridge oscillates around our predicted spiral.
The deviations become larger where the spiral rolls-up (\cref{fig:FTLEregimes}b).
Finally, at $\Rey = \num{8380}$ the shear layer is no longer visible in the vorticity field snapshot and we observe discrete secondary vortices instead (\cref{fig_Kadencomparison}c).
The wavelength of the nFTLE ridge fluctuations has decrease with the increase of the Reynolds number towards the discrete shedding regime.
The spiral computed with \cref{eq:modKaden} represents the middle line along which the FTLE ridge oscillates (\cref{fig:FTLEregimes}b).
We can distinguish four lobes on the outside of the predicted spiral that surround four secondary vortices in \cref{fig_Kadencomparison}c.
With increasing value of the Reynolds number, we can distinguish three regimes: a first regime ($\Rey < \num{1500}$) which is characterised by a stable shear layer, a transitional regime ($\num{1500}<\Rey<\num{4000}$) which is characterised by first signs of instability, and a discrete vortex shedding regime ($\Rey> 4000$) where vorticity is only observed in isolated patches.
For the three \Rey regimes observed, the modified Kaden spiral is able to predict the roll-up of the shear layer and the path of the secondary vortices for the entire rotation of the plate.

\begin{figure}
\begin{center}\includegraphics{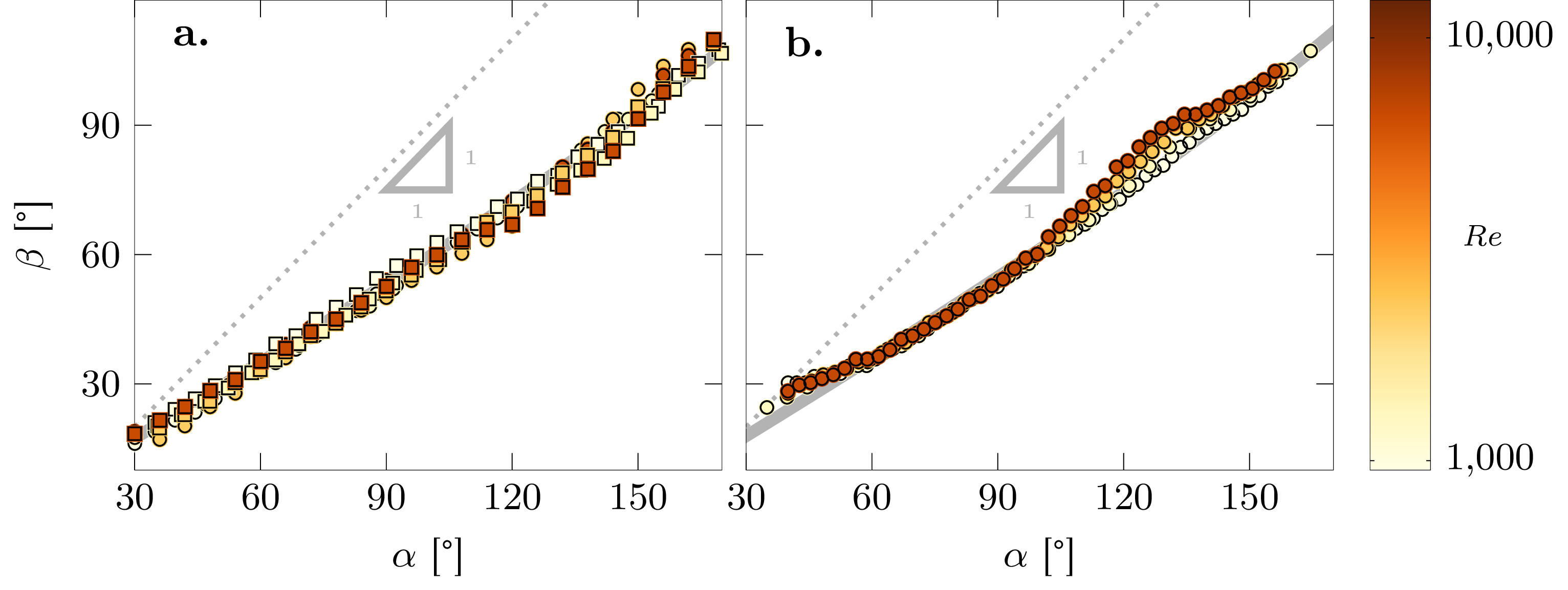}\end{center}
\caption{Variation of the angular location of the primary vortex ($\beta$) with convective time indicated by the plate's rotation angle ($\alpha$) for (a) rotations around the mid-length and rotations around the edge, both with $c=\SI{4}{\centi\meter}$ ; and (b) rotation around the mid-length with  $c=\SI{6}{\centi\meter}$.}
\label{fig:Betaintime}
\end{figure}

To further quantitatively validate our modified Kaden spiral model, we compare the measured angular locations of the primary vortex as a function of the convective time $\alpha$ with the predicted model results in \cref{fig:Betaintime} for different \Rey.
The angular position $\beta$ of the primary vortex increases with $\alpha$.
The relationship between $\beta$ and $\alpha$ is close to, but not entirely linear.
The trajectory of the primary vortex is completely independent of the Reynolds number and is accurately predicted by the modified Kaden spiral.
The trajectory is also not influenced by the total length of the plate.
The measured data presented in \cref{fig:Betaintime}a include results from the plates with the rotational location at the mid-length and from the plates with the rotational location at one end of the plate.
The distance between the rotational point and the tip is the same in both cases.
From the perspective of vortex formation and shear layer roll-up, a plate with a length of $\SI{4}{\centi\meter}$ that rotates around one end is equivalent to a $\SI{8}{\centi\meter}$ long plate rotating around its centre location.
The presence of a flipped and mirrored vortex system and shear layer topology on the other side of the longer plate has no influence on the roll-up nor on the trajectory of the primary vortex for the plate geometries and Reynolds numbers tested here.

The influence of the distance between the rotational point and the plate tip, referred to as the chord length here, is analysed by considering a plate with length $\SI{12}{\centi\meter}$ and chord length $\SI{6}{\centi\meter}$.
For rotational motions with the longer plate, we observe the same shear layer topology for the same \Rey-regimes described before.
The modified Kaden spiral predictions still provide an excellent prediction of the shear layer roll-up and the trajectory of the primary vortex in \cref{fig:Betaintime}b.
The angular velocity in terms of $\dd \beta/\dd \alpha$ is slightly increased for the higher chord length plates and a higher value of $\eta=\num{1.59e-2}$ was used for the modified Kaden spiral predictions of the larger chord length wing.
For the two different chord lengths, the ratio $\eta/c=0.260\pm0.005$.
To take into account the influence of the chord length in our modified Kaden spiral model, we replace the empirical constant $\eta$ in \cref{eq:modKaden} with $\eta' c$ to obtain:
\begin{equation}\label{eq:modmodKaden}
r=\eta' c \alpha \left(\frac{\alpha}{\theta}\right)^{2/3},
\end{equation}
where $\eta'=0.260$ for all data presented in this paper.

\begin{figure}
\begin{center}\includegraphics{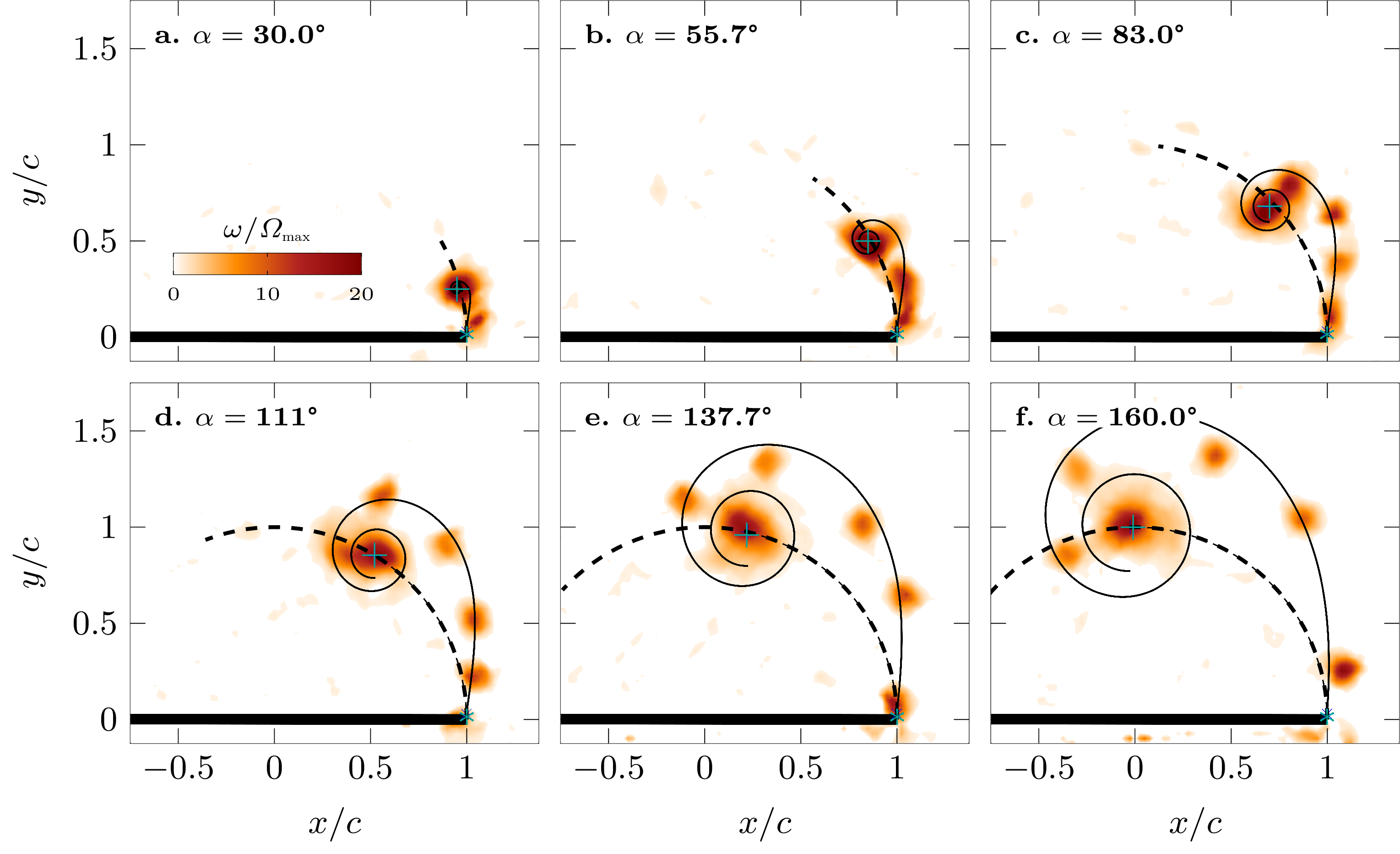}\end{center}
\caption{
Temporal evolution of secondary vortices at different angular positions (a) $\alpha=\ang{30.0}$, (b) $\alpha=\ang{55.7}$, (c) $\alpha=\ang{83.0}$, (d) $\alpha=\ang{110.5}$, (e) $\alpha=\ang{137.7}$, and (f) $\alpha=\ang{160.0}$ for $\Rey = \num{8380}$.
The black curve is the modified Kaden's spiral, whose centre and end are the primary vortex centre ($+$) and the right top plate edge ($\ast$).
The spiral is only plotted for $\theta$ ranging from \kindex{\theta}{tip} to $4\pi$.
The dashed line represents the plate tip trajectory.
}
\label{fig:spiral_evolution}
\end{figure}

\subsection{Timing of the secondary vortex shedding}
In the next part, we focus our attention on the successive shedding of secondary vortices.
The first step is to determine if these secondary vortices are generated from the stretching of an initially unstable shear layer or if they are discretely released after the separation of the primary vortex.
\Cref{fig:spiral_evolution} shows the flow topology at different plate angular positions for $\Rey = \num{8380}$.
Between $\alpha=\ang{0}$ and $\alpha=\ang{30}$ the primary vortex centre is close to the plate tip and no secondary vortices are observed.
At $\alpha=\ang{30}$, the primary vortex has moved away from the tip along the circular tip trajectory and a first secondary vortex forms (\cref{fig:spiral_evolution}a).
The first secondary vortex drifts towards the primary vortex core and they merge as a consequence of their mutual interaction (\cref{fig:spiral_evolution}b).
The formation and shedding of successive secondary vortices is repeated along the entire motion.
Each vortex is independently formed and subsequently released from the plate tip.
In this situation, the shear layer appears as a cloud of vorticity close to the plate tip from which vortices are discretely detached.
Once the secondary vortices shed, they move away and are located along the modified time-varying Kaden spiral (\cref{fig:spiral_evolution}c-e).
Vortices closer to the primary vortex deviate slightly from the predicted spiralling curve only when the plate rotation is about to finish (\cref{fig:spiral_evolution}f).

\begin{figure}
\begin{center}\includegraphics{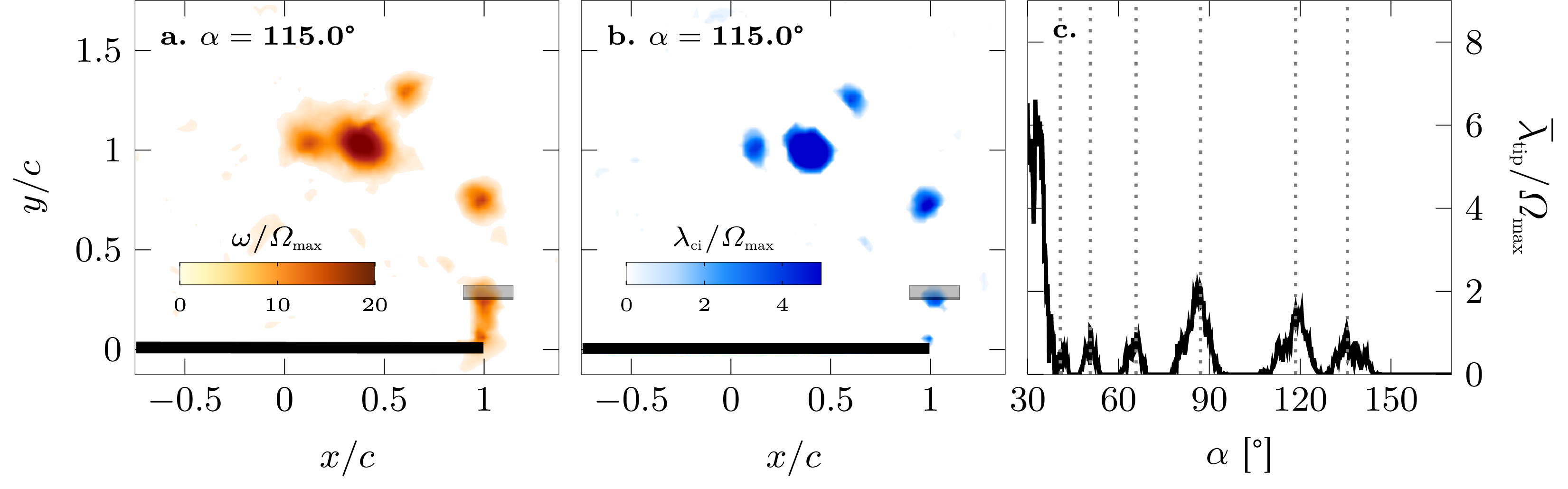}\end{center}
\caption{
Snapshot of the (a) vorticity field and (b) the swirling strength at $\alpha=\ang{115}$ for $\Rey = \num{8380}$.
The black rectangle corresponds to the region in which $\kindex{\overline{\lambda}}{tip}$ is computed.
(c) Evolution of $\kindex{\overline{\lambda}}{tip}$ as a function of the angular position of the plate.
The dotted lines mark the local maxima in the average tip swirling strength.
The timing of the local maxima are related to the separation angle of subsequent secondary vortices.
}
\label{fig:peaks_curve}
\end{figure}

The second step is to compute the timing of secondary vortices.
If we consider the vorticity field, the constant presence of the cloud of vorticity close to the tip hampers the identification of the separation time.
To estimate the timing of shedding of the individual vortices we use the swirling strength criterion by \citet{zhouMechanismsGeneratingCoherent1999}.
A vortex is considered a connected region where the value of the swirling strength $\kindex{\lambda}{ci}$ is positive.
The swirling strength criteria allows us to distinguish more reliably whether a region of high vorticity concentration indicates the presence of a secondary vortex or whether it is due to a strong shear flow (see \cref{fig:peaks_curve}a,b).
To determine the timing of release of subsequent secondary vortices, we calculate and analysed the evolution of the local average swirling strength, denoted by $\kindex{\overline{\lambda}}{tip}$, in a small rectangular region close to the tip of the plate.
As we are purely interested in the counterclockwise rotating structure here, we only count the positive swirling strength in regions where the vorticity is positive.
The location of the probing region is indicated in \cref{fig:peaks_curve}a,b and an example of the resulting temporal evolution of the local average tip swirling strength for $\Rey=\num{8380}$ is presented in \cref{fig:peaks_curve}c.
The temporal evolution of $\kindex{\overline{\lambda}}{tip}$ has a global maximum and first peak at $\alpha = \ang{32.6}$ which is followed by six clearly distinguishable smaller peaks.
The initial peak corresponds to the shedding of the primary vortex, and the subsequent smaller peaks mark the shedding of individual secondary vortices.
The average swirling strength systematically drops to zero in between the individual peaks, further supporting the conclusion that the secondary vortices are discretely released from the tip of the plate.
The timing of the local maxima of $\kindex{\overline{\lambda}}{tip}$ is used to further analyse the shedding timing of the secondary vortices.
This strategy to determine the timing of secondary vortex shedding is simple yet robust and allows for a systematic and automated extraction of the timings for all measurements.
The results depend slightly on the location and size of the probing region which were carefully selected based on a sensitivity analysis (\cref{app:sensitivityanal}).

\begin{figure}
\begin{center}\includegraphics{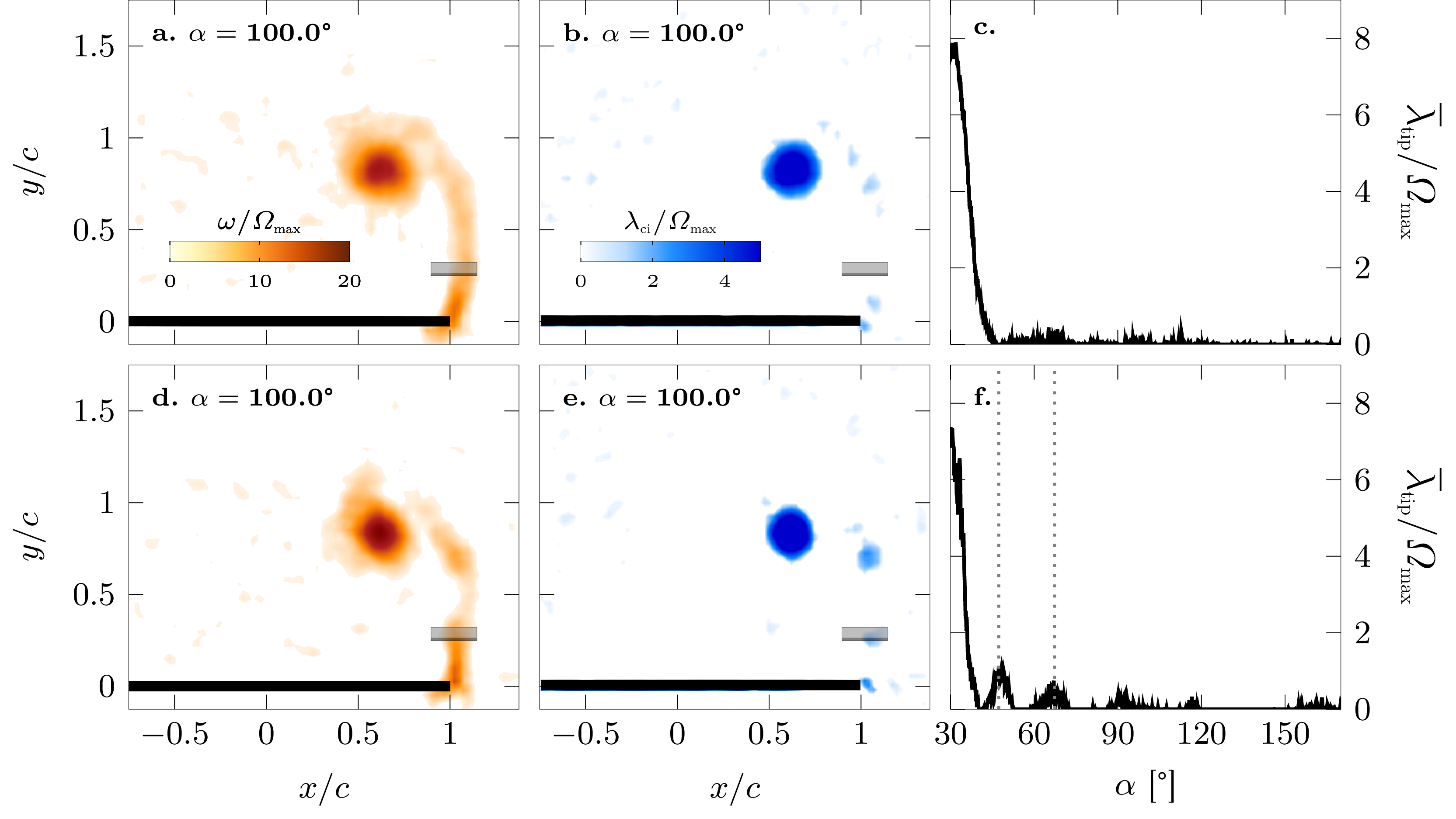}\end{center}
\caption{
Snapshots of the (a,d) vorticity field and (b,e) swirling strength at $\alpha=\ang{100}$ and (c,f) evolution of $\kindex{\overline{\lambda}}{tip}$ as a function of the angular position of the plate.
The first row corresponds to $\Rey = \num{840}$ at which the shear layer appears continuous.
The second row is for $\Rey = \num{1955}$ at which the shear layer shows signs of instability.
}
\label{fig:newcomparison}
\end{figure}

Results of the timing extraction strategy for $\Rey = \num{840}$ and $\Rey = \num{1955}$ are summarised in \cref{fig:newcomparison}.
For the lowest Reynolds number $\Rey = \num{840}$, we have a continuous stable shear layer and the associated snapshot of the swirling strength at $\alpha=\ang{100}$ in \cref{fig:newcomparison}a shows a single isolated coherent structure and no sign of secondary vortices.
This is confirmed by the time evolution of $\kindex{\overline{\lambda}}{tip}$ (\cref{fig:newcomparison}a) that exhibits a single peak at $\alpha=\ang{31.9}$.
No other peaks are observed afterwards confirming that the shear layer is a continuous layer of fluid without the presence of any instabilities for this Reynolds number.
For the intermediate Reynolds number $\Rey = \num{1955}$, the shear layer topology appeared to be undulating with some localised concentrations of high vorticity along it (\cref{fig_Kadencomparison}b).
The temporal evolution of the average tip swirling strength reveals the shedding of two secondary coherent structures formed after the primary vortex (\cref{fig:newcomparison}b).
These two structures are formed and released from the tip and they are not formed afterwards due to the stretching of the shear layer which does not become clear based solely on the vorticity flow topology.

\begin{figure}
\begin{center}\includegraphics{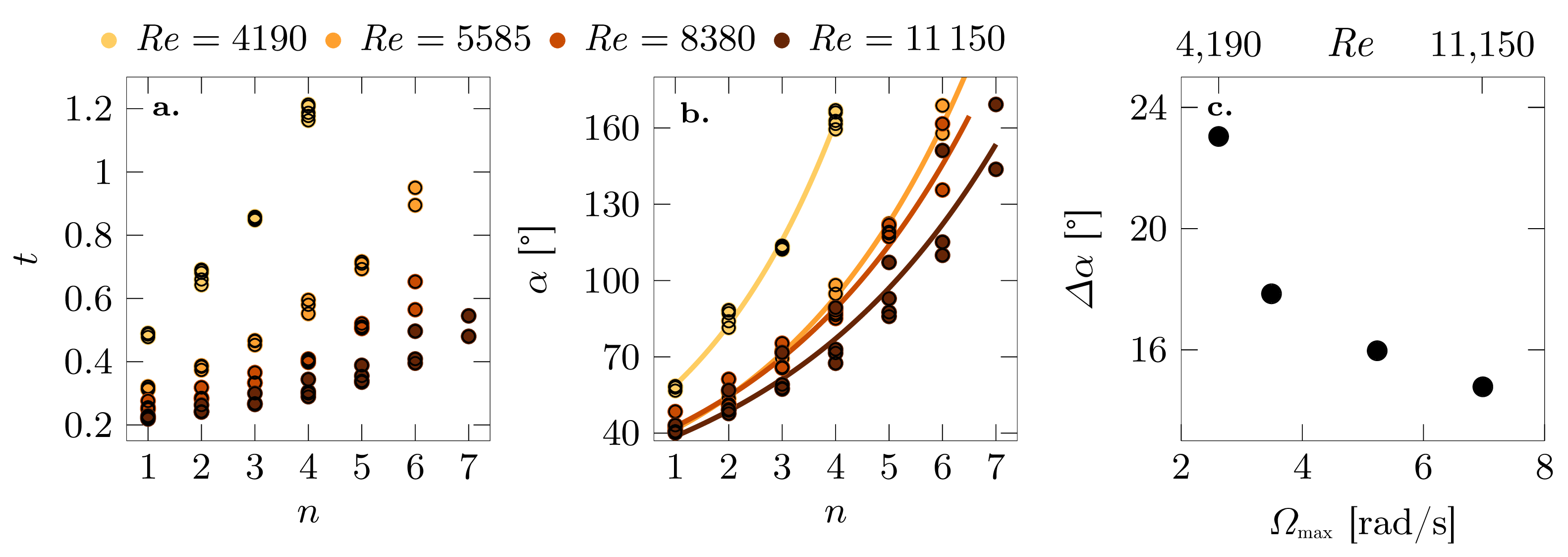}\end{center}
\caption{
Delay between the successive shedding of secondary vortices in terms of (a) dimensional time and (b) convective time or angular distance between secondary vortices as of the shedding order $n$.
The solid lines are the fit of the angular distance between vortices and $n$.
(c) Coefficient $\Delta\alpha$ as a function of the maximum rotational speed of the plate.
}
\label{fig:newtiming}
\end{figure}

The experiments are repeated five times at each Reynolds number.
The separation time and angle of successive secondary vortices are computed and analysed for all experiments with $\Rey>4000$ corresponding to the discrete shedding regime.
The timing of the secondary vortex shedding versus the number corresponding to the order of successive shedding is presented in \cref{fig:newtiming}.
In general, the time interval between successive vortices increases the more secondary vortices have been released and the time interval decreases with increasing Reynolds number, yielding a larger total number of secondary vortices at the end of the \ang{180} plate rotation.
If we hypothesise that the strength of the secondary vortices remains approximately constant, which is confirmed by the experimental data, then the increase in time interval should be due to a decrease in the circulation feeding rate by the shear layer.
This feeding rate is related to the shear rate of at the tip of the plate and can be estimated by:
\begin{align}
\frac{\dd \Gamma}{\dd t}&\propto\frac{\kindex{v}{out}^2-\kindex{v}{in}^2}{2}\approx\frac{(\Omega c/2)^2 - \kindex{v}{in}^2(t)}{2}
\end{align}
where \kindex{v}{out} refers to the velocity at the outer side of the shear layer, which equals the tip velocity $\Omega c/2$ and \kindex{v}{in} refers to the velocity at the inner side of the shear layer.
The velocity at the inner side \kindex{v}{in} is close to zero during the initial part of the rotation as the plate rotates in a quiescent fluid and increases due to the accumulation of vortex induced velocity components along the direction of the plate's motion.
The feeding rate thus decreases when the rotational velocity and the Reynolds number decrease and when the induced velocity due to an increased number of released vortices.
This explains the general trends observed in \cref{fig:newtiming}a,b.

To quantify the evolution of the shedding timing of the secondary vortices, we fit the measured values in \cref{fig:newtiming}b with a power law in the form:
\begin{equation}\label{eq:feedingrate}
\alpha(n) = \kindex{\alpha}{0}(1+\Delta\alpha)^n
\end{equation}
where \kindex{\alpha}{0} and $\Delta\alpha$ are fitting constants and $n$ counts the number of secondary vortices.
This power law is suitable to represent the timing dynamics for all Reynolds numbers as all fits have a $R^2$-value above $\SI{99}{\percent}$.
The fitting parameter $\Delta\alpha$ indicates the relative increase in $\alpha$ between the convective timing of successive secondary vortices, i.e. $\kindex{\alpha}{n+1}/\kindex{\alpha}{n}=1+\Delta\alpha$.
A higher value of $\Delta\alpha$ indicates a larger delay between successive vortices and a lower total number of vortices shed at the end of the motion.
The evolution of $\Delta\alpha$ as a function of the maximum rotational speed is presented in \cref{fig:newtiming}c.
The results confirm the general decrease in the time delay between the release of subsequent vortices for increased rotational velocity of the plate which yields an increased feeding rate according to \cref{eq:feedingrate}.
\section{Conclusion}
The roll-up of a shear layer behind a rotating plate in a quiescent fluid is experimentally studied for different rotational velocities or Reynolds numbers.
Particular focus was directed towards the formation, trajectory, and timing of secondary vortices.

Based on the time-resolved PIV, we identified three Reynolds number regimes based on the stability of the shear layer.
For $\Rey < \num{1500}$, a stable shear layer in the form of a continuous band of vorticity is observed that rolls up into a single coherent primary vortex.
For $\Rey > \num{4000}$, the shear layer is unstable and secondary vortices are discretely released from the plate's tip during the rotation.
In the intermediate regime for $\num{1500} <\Rey < \num{4000}$, first signs of instability appear.
The shear layer is still a continuous band of vorticity but it shape is wavier and localised concentrations of higher vorticity emerge.
In all three regimes, the centre of the primary vortex is located on the plate tip trajectory and the shear layer topology matches a spiral shape similar to the roll up of a free shear layer.
A modified version of the Kaden spiral is proposed to describe and predict the temporal evolution of the shear layer roll up.
The key modification is the replacement of the constant dimensional Kaden constant $K$ by a factor $\eta'c\alpha$ that increases linearly with the rotational angle of the plate and takes into the effect of the chord length.
A single value of $\eta'$ has been empirically determined for all experimental conditions presented in this paper.
The proposed modified Kaden spiral model describes the spatiotemporal evolution of the shear layer and accurately predicts the trajectory of the centre of the primary vortex for all Reynolds numbers and different plate dimensions.

The timing of secondary vortices shedding for Reynolds numbers in the discrete shedding regimes is determined using the swirling strength criterion.
The swirling strength fields confirm that secondary vortices form directly at the tip of the plate and not further downstream due to the stretching of the shear layer.
The separation time of each secondary vortex is identified as a local maximum in the temporal evolution of the average swirling strength close to the plate tip.

The time interval between the release of successive vortices is not constant during the rotation but increases the more secondary vortices have been released.
The shedding time interval also increases with decreasing Reynolds number, yielding a lower total number of secondary vortices at the end of the \ang{180} plate rotation for lower \Rey.
The increased time interval under both conditions is due to a reduced circulation feeding rate.

\appendix
\section{Derivation of the modified Kaden spiral}\label{app:modKaden}

\begin{figure}
\begin{center}
\includegraphics{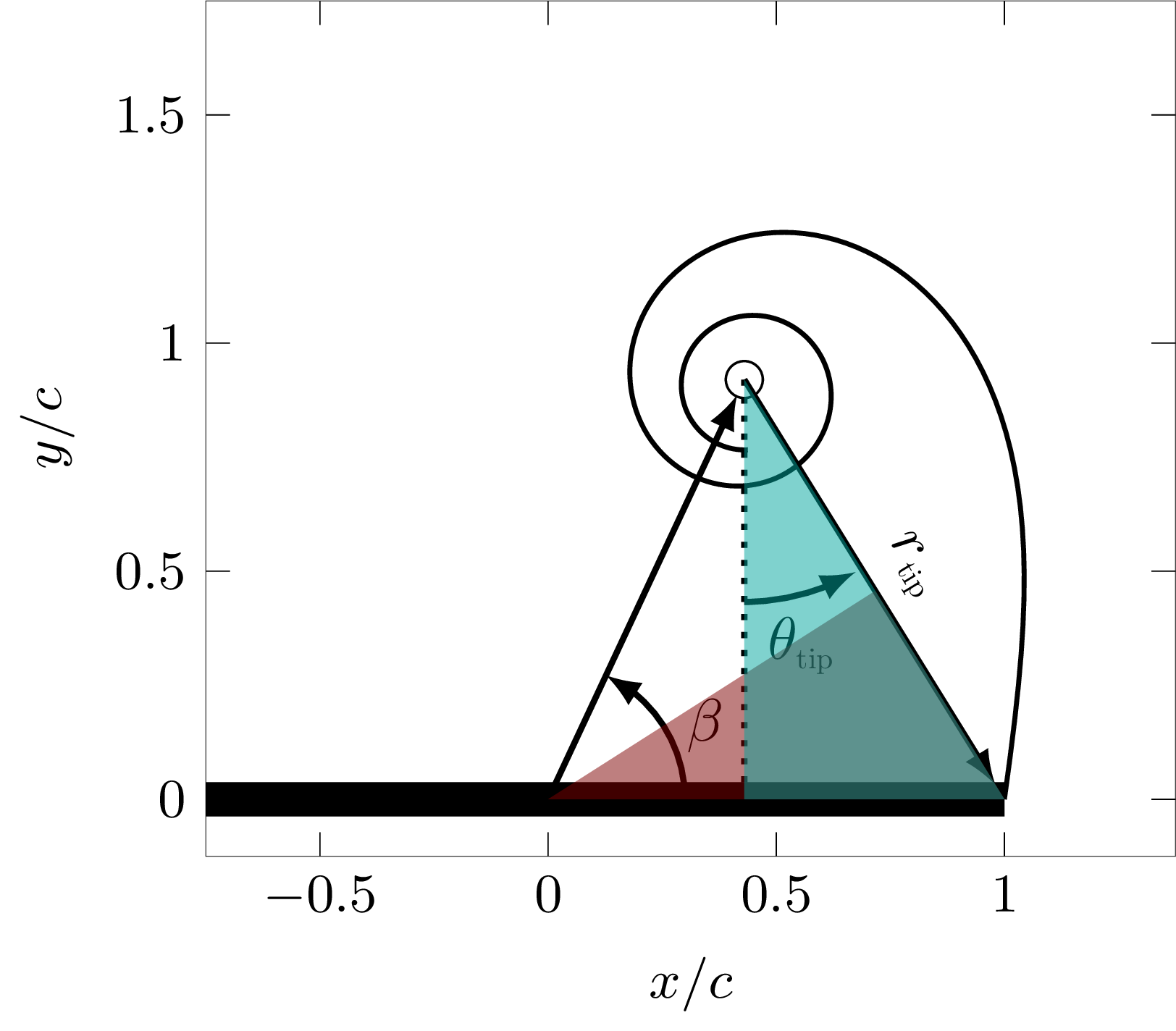}
\end{center}
\caption{Definition of the radial and angular spiral coordinates and its orientation with respect to the plate's frame of reference.
The trigonometric relationships \cref{eq:beta1} and \cref{eq:beta2} are obtained in the shaded triangles.
}\label{fig:appgeom}
\end{figure}

The modified version of the Kaden spiral we propose takes into account the temporal increase in the distance between the primary vortex and the tip of the plate where the bound end of the vortex sheet is fixed.
The angular coordinate along the spiral that marks the bound end of the vortex sheet is denoted by \kindex{\theta}{tip}.
The value of \kindex{\theta}{tip} is determined at every time step based solely on the observation that the primary vortex moves along a path that matches the plate tip trajectory.
Based on this purely geometric constraint, we also directly obtain the radial spiral coordinate where the modified Kaden spiral meets the plate tip, indicated by \kindex{r}{tip}, and the angular location of the primary vortex with respect to the plate, denoted by $\beta$.
Their detailed derivation is given here.

We consider the flow situation after the plate has rotated for a given $\alpha$ in the plates frame of reference in \cref{fig:appgeom}.
The plate tips trajectory is indicated by the dashed line.
The primary vortex is located on that circular trajectory.
Its angular position with respect to the plate's centre of rotation and tip is indicated by $\beta$.

Consider that we have shifted the modified spiral defined by \cref{eq:modmodKaden} such that the spiral centre ($r=0$, $\theta\rightarrow \infty$) is located in the centre of the primary vortex.
The radial and angular location of the plate tip in the spiral coordinates are given by (\kindex{r}{tip}, \kindex{\theta}{tip}) as indicated in \cref{fig:appgeom}.
For a given spiral form, there is only one solution for $\beta$ that allows the spiral to go through the plate tip.
This solution can be found by ensuring that the trigonometric relationships for the two triangles outlined in \cref{fig:appgeom}b are met:
\begin{align}
\beta &= 2\arcsin (\kindex{r}{tip}\cos \kindex{\theta}{tip}/2c)\label{eq:beta1}\\
\beta &= \arcsin(\kindex{r}{tip}\cos \kindex{\theta}{tip}/c)\label{eq:beta2}\quad.
\end{align}
The distance between the primary vortex centre and the tip of the plate, $\kindex{r}{tip}$, is determined through the modified Kaden spiral (\cref{eq:modmodKaden}), for $\theta = \kindex{\theta}{tip}$.
In this way, \cref{eq:beta1} and \cref{eq:beta2} are only functions of \kindex{\theta}{tip}, which is computed by equalising the two relationships.
For $\beta > \pi/2$, we need to use
\begin{equation}\label{eq:beta3}
	\beta = \pi - \arcsin(\kindex{r}{tip}\cos \kindex{\theta}{tip}/c)
\end{equation}
instead of \cref{eq:beta2}.
Once \kindex{\theta}{tip} is retrieved, we substitute it into \cref{eq:beta1} to obtain the angular position $\beta$ of the primary vortex.
From the value of $\beta$, we compute the cartesian coordinates of the primary vortex centre, which corresponds to the centre of our predicted spiral model.
The full spiral is finally obtained for every plate angular position $\alpha$, using \cref{eq:modKadenspiral} with $\theta\in[\kindex{\theta}{tip}, \infty]$.

\section{Sensitivity analysis of the location and size of the average tip swirling strength probing region}\label{app:sensitivityanal}

The local average swirling strength $\kindex{\overline{\lambda}}{tip}$ reaches a local maximum value when most of the vortex fills the selected rectangular region.
If the position and dimension of the rectangular region is not properly set, the identification of the separation time through local peaks loses accuracy.
We perform a sensitivity analysis of the best position and dimension of the rectangular region.
The first thing to set is the centre of the rectangle.
We observed that when the core centre of a secondary vortex is approximately $\SI{1}{\centi\meter}$ above the plate tip, the following secondary vortex starts growing.
Since the trajectory of each secondary vortex is predicted by the modified Kaden's spiral (\cref{eq:modmodKaden}), we decide to place the centre of the rectangle along the spiral, $\SI{1}{\centi\meter}$ above the tip.
The area of the rectangle has to be big enough to fully include the vortex but it does not have to include the swirling strength associated to the plate and to the other secondary vortices.
For the plate chord $c = \SI{4}{\centi\meter}$, a rectangle base of $0.25c$ is sufficiently large to include the radial dimension of the secondary vortex, excluding the swirling strength of the plate.
A height of $0.15c$ allows to have one secondary vortex at a time in the selected rectangle.

\section*{Declaration of interest}
The authors report no conflict of interest.

\bibliographystyle{jfm}
\bibliography{p2v3}
\end{document}